\documentclass[showpacs,preprintnumbers,amsmath,amssymb]{revtex4}

\usepackage{graphicx}
\usepackage{dcolumn}
\usepackage{bm}

\begin{document}

\title{Remote Preparation of Mixed States via Noisy Entanglement}

\author{Guo-Yong Xiang\thanks{%
Email address: gyxiang@mail.ustc.edu.cn}, Jian Li, Bo Yu, and
Guang-Can
Guo\thanks{%
Email address: gcguo@ustc.edu.cn}}
\address{Key Laboratory of Quantum Information, University of Science and Technology\\
of China, CAS, Hefei 230026, People's Republic of
China\bigskip\bigskip }

\begin{abstract}
\ We present a practical and general scheme of remote preparation
for pure and mixed state, which is proposed with an auxiliary
qubit and controlled-NOT gate. We discuss the remote state
preparation (RSP) in two important types of decoherent channel
(depolarizing and dephasing). We realize RSP in the dephasing
channel in our experiment by using spontaneous parametric down
conversion (SPDC), linear optical elements and single photon
detector. Our experiment results match the theoretical prediction
well.
\end{abstract}

\maketitle

\section{Introduction}
A principal goal of quantum information theory is to understand
the
resources necessary and sufficient for intact transmission of quantum states%
\cite{Nielsen}. Obviously, Alice (the sender) can either
physically send the particle (which is not interesting) or send a
double infinity of bits of
information across a classical channel to Bob. However, quantum teleportation%
\cite{Bennett93} has provided us an interesting way to
transmitting an arbitary quantum state using one maximal-entangled
state (1 ebit) and two classical bits information (2 cbits).
Recently, remote state preparation
(RSP)\cite{Lo00,Pati00,Ben01,Dev01,Ber03,Leu03,Peng02,Pati02,Ye03,Ben03}
attracts attention of many scientists. RSP provides us a simple
method to transmit pure quantum states using entanglement and
classical communication when the sender knows the transmitted
state and the receiver has partial knowledge of it. RSP protocols
more economical than teleportation were found for certain
ensembles\cite {Lo00,Pati00,Ben01,Dev01,Ber03,Leu03,Pati02,Ben03}.
The resources of 1 ebit and 2 cbits are sufficient and necessary
for teleportation, while it can be reduced to 1 ebit and 1 cbit in
asymptotics for remote state preparation. Such trade-off has been
extent among resources of noiseless classical channels, noiseless
quantum channels and maximally-entangled states in the generalized
remote state preparation\cite{Abey03}. An experiment of RSP has
been demonstrated in nuclear magnetic resonance (NMR)
systems\cite{Peng02}.

Recently, the RSP protocol is generalized from pure states to
mixed ones. The essence is to replace von Neumann measurement by
positive operator-valued measure (POVM)\cite{Ber03}. Here we
present a scheme for remote preparation of mixed states (including
pure ones) and realize it for the polarization states of single
photons.

All protocols about RSP above is (deterministic or probabilistic)
exact, in which state that Bob gets is as same as the one Alice
wants to prepare. It is based on the maximally-entangled state
shared between the two sides. However, entanglement is fragile
under the interaction with environment. In this paper, two types
of decoherence, depolarizing and dephasing, are discussed by using
the state fidelity in theory as well as in our experiment.

This paper is organized as following. A RSP scheme of mixed states
is presented in section II. The effect of decoherence is discussed
with state fidelity in section III. In section IV, we carry out an
experiment to remotely prepare pure and mixed states with
dephaseing noisy entanglement. And a conclusion will be given in
section V.

\section{Remote preparation of mixed states}

An arbitrary state of single qubit can be represented as a vector $%
\overrightarrow{r}$ on or inside the Bloch sphere (see Fig.1)
\begin{equation}
\rho (\overrightarrow{r})=\frac 12\left( I+\overrightarrow{r}\cdot
\overrightarrow{\sigma }\right)
\end{equation}
where $\overrightarrow{r}=\left( r\sin \theta \cos \phi ,r\sin
\theta \sin
\phi ,r\cos \theta \right) $ and $0\leq r\leq 1$, $0\leq \theta \leq \pi $, $%
0\leq \phi \leq 2\pi $. The state is a pure one $\left| \psi
\right\rangle =\cos \frac \theta 2\left| 0\right\rangle +e^{i\phi
}\sin \frac \theta 2\left| 1\right\rangle $ for $r=1$ (on the
sphere), or a mixed one for $r<1$ (inside the sphere). Specially,
it is a maximally-mixed one $\frac 12\left( \left| 0\right\rangle
\left\langle 0\right| +\left| 1\right\rangle \left\langle 1\right|
\right) $ for $r=0$ (zero vector).

In RSP protocol for pure states, a state subset $\chi $ is
pre-agreed by Alice and Bob. Particles $A$ and $B$ , shared by
Alice and Bob, are in maximally entangled state
\begin{equation}
\left| \Psi ^{-}\right\rangle _{AB}=\frac 1{\sqrt{2}}\left( \left|
0\right\rangle _A\left| 1\right\rangle _B-\left| 1\right\rangle
_A\left| 0\right\rangle _B\right) ,
\end{equation}
and Alice wants to help Bob prepare a state $\left| \psi
\right\rangle \in
\chi $ in the distance. Here $\left| \psi \right\rangle $ is selected from $%
\chi $ randomly by Alice and is unknown to Bob. For example, $\chi
$ can be the equatorial or polar great circle on the Bloch sphere
and $\left| \psi \right\rangle $\ is randomly selected from it.

To complete RSP, we can expand $\left| \Psi ^{-}\right\rangle
_{AB}$ in the basis \{$\left| \psi \right\rangle ,\left| \psi
_{\perp }\right\rangle $\},
\begin{equation}
\left| \Psi ^{-}\right\rangle _{AB}=\frac 1{\sqrt{2}}\left( \left|
\psi \right\rangle _A\left| \psi _{\perp }\right\rangle _B-\left|
\psi _{\perp }\right\rangle _A\left| \psi \right\rangle _B\right)
\end{equation}
where $\left| \psi _{\perp }\right\rangle =e^{i\phi }\sin \frac
\theta 2\left| 0\right\rangle -\cos \frac \theta 2\left|
1\right\rangle $. Alice would perform a von Neumann measurement
$\left\{ \left| \psi \right\rangle ,\left| \psi _{\perp
}\right\rangle \right\} $ on her particle $A$ and send the result
$0/1$ ($\left| \psi \right\rangle /\left| \psi _{\perp
}\right\rangle $) to Bob. Or she can ($i$) perform a unitary rotation $%
U(\theta ,\phi )^{\dagger }$ on $A$,
\begin{equation}
U_A(\theta ,\phi )^{\dagger }\left| \Psi ^{-}\right\rangle _{AB}=\frac 1{%
\sqrt{2}}\left( \left| 0\right\rangle _A\left| \psi _{\perp
}\right\rangle _B-\left| 1\right\rangle _B\left| \psi
\right\rangle _B\right)
\end{equation}
where
\begin{equation}
U(\theta ,\phi )=\left(
\begin{array}{cc}
\cos \theta /2 & -e^{-i\phi }\sin \theta /2 \\
e^{i\phi }\sin \theta /2 & \cos \theta /2
\end{array}
\right) ;
\end{equation}
($ii$) carry out a von Neumann measurement $\{\left|
0\right\rangle ,\left| 1\right\rangle \}$; and ($iii$) send the
result $0/1$ ($\left| 0\right\rangle /\left| 1\right\rangle $) to
Bob (see Fig.2a).\ If the
result is $1$, Bob will find his particle $B$ in $\left| \psi \right\rangle $%
, which is the state Alice wants to prepare. For the result $0$,
$B$ will be in $\left| \psi _{\perp }\right\rangle $\ and an
operation \ is need for Bob to flip $\left| \psi _{\perp
}\right\rangle $\ into $\left| \psi \right\rangle $. Generally,
such operation is unavailable because the universal $NOT$ gate is
forbidden for Bob has no knowledge of the state. However, it is
possible for some special $\chi $, such as $\sigma _Z$ for the
polar greatest circle and $i\sigma _Y$ for the equatorial circle.
In the rest of this paper, we only discuss the situation that
Alice obtain the result $1$.

In \cite{Ber03}, Berry and Sanders have generalized the protocol
to mixed state preparation by using POVM instead of von Neumann
measurement. The mixed state to be prepared (see eq.(1) ) can be
decomposed into
\begin{equation}
\rho (\overrightarrow{r})=\frac{1+r}2\left| \psi \right\rangle
\left\langle \psi \right| +\frac{1-r}2\left| \psi _{\perp
}\right\rangle \left\langle \psi _{\perp }\right|
\end{equation}
To complete RSP of $\rho (\overrightarrow{r})$, Alice would carry
out a two-element POVM $\{\Pi ^0,\Pi ^1\}$ instead of $\{\left|
0\right\rangle ,\left| 1\right\rangle \}$,
\begin{equation}
\Pi ^1=\frac{1-r}2\left| 0\right\rangle \left\langle 0\right| +\frac{1+r}%
2\left| 1\right\rangle \left\langle 1\right| ,\Pi ^0=I-\Pi ^1.
\end{equation}
If Alice gets the result 1, $B$ will be in $\rho
(\overrightarrow{r})$. The essence is the realization of POVM.
Here, the POVM can be performed by an auxiliary qubit and
$Controlled-NOT$ operation. The whole process of RSP can be
divided into the following five steps (see Fig.2b). $1$) a unitary
rotation $U(\theta ,\phi )^{\dagger }$ on $A$; $2$) a
$Controlled-NOT$ operation $U_{CNOT}$ where $A$ is the controller
and the auxiliary qubit $a$ in initial state $\left|
0\right\rangle $ is the target; $3$) another unitary rotation
$U_A(r)$ on $A$; $4$) a von Neumann measurement on $A$ ; and $5$)
sending the result to Bob. And it can be represented as
\begin{eqnarray}
&&\left| \Psi ^{-}\right\rangle _{AB}\left| 0\right\rangle _a\stackrel{%
U_A(\theta ,\phi )^{\dagger }}{\rightarrow }\frac
1{\sqrt{2}}\left( \left| 0\right\rangle _A\left| \psi _{\perp
}\right\rangle _B-\left| 1\right\rangle
_B\left| \psi \right\rangle _B\right) \left| 0\right\rangle _a  \nonumber \\
&&\stackrel{U_{CNOT}(A\rightarrow a)}{\rightarrow }\frac
1{\sqrt{2}}\left( \left| 0\right\rangle _A\left| \psi _{\perp
}\right\rangle _B\left| 0\right\rangle _a-\left| 1\right\rangle
_B\left| \psi \right\rangle _B\left|
1\right\rangle _a\right)  \nonumber \\
\stackrel{U_A(r)}{\rightarrow }\left| \varphi \right\rangle _{ABa}
&=&\frac 12(\sqrt{1+r}\left| 0\right\rangle _A\left| \psi _{\perp
}\right\rangle _B\left| 0\right\rangle _a+\sqrt{1-r}\left|
0\right\rangle _A\left| \psi
\right\rangle _B\left| 1\right\rangle _B \\
&&+\sqrt{1-r}\left| 1\right\rangle _A\left| \psi _{\perp
}\right\rangle _B\left| 0\right\rangle _a-\sqrt{1+r}\left|
1\right\rangle _A\left| \psi \right\rangle _B\left| 1\right\rangle
_a  \nonumber),
\end{eqnarray}
where
\begin{equation}
U(r)=\left(
\begin{array}{cc}
\sqrt{\frac{1+r}2} & -\sqrt{\frac{1-r}2} \\
\sqrt{\frac{1-r}2} & \sqrt{\frac{1+r}2}
\end{array}
\right) .
\end{equation}
For the result $1$, $B$ will be in
\begin{equation}
\rho _B=\frac{Tr_{Aa}[\left| 1\right\rangle _{A\ A}\left\langle
1\right| \left| \varphi \right\rangle _{ABa\ ABa}\left\langle
\varphi \right| ]}{ Tr_{ABa}[\left| 1\right\rangle _{A\
A}\left\langle 1\right| \left| \varphi \right\rangle _{ABa\
ABa}\left\langle \varphi \right| ]}.
\end{equation}
It's easy to see that the protocol is valid for pure state by
setting $r=1$.

\section{Effects of noisy entanglement}

All the discussions above, both pure and mixed states, are based
on the maximally entangled states shared between Alice and Bob.
Due to the interaction with the environment, the entanglement will
be partially destroyed. Such decoherence is possible during the
distribution or storage of entanglement, So the final result of
RSP will be influenced.

Generally, all physical processes, including decoherence
evolution, can be represented by a complete positive map. Suppose
Alice and Bob share an entangled state after decoherence,
\begin{equation}
\rho _{AB}=\widehat{S}\left( \left| \Psi ^{-}\right\rangle _{AB\
AB}\left\langle \Psi ^{-}\right| \right) ,
\end{equation}
where $\widehat{S}$ is the operator of decoherence evolution. The
final state that Bob obtains after the five steps above ($1-5$)
will be
\begin{equation}
\rho _B=\frac{Tr_{Aa}[\left| 1\right\rangle _{A\ A}\left\langle
1\right| \varrho _{ABa}]}{Tr_{ABa}[\left| 1\right\rangle _{A\
A}\left\langle 1\right| \varrho _{ABa}]},
\end{equation}
where
\begin{equation}
\varrho _{ABa}=U_A(r)\otimes U_{CNOT}(A:a)\otimes U_A(\theta ,\phi
)^{\dagger }\rho _{AB}U_A(\theta ,\phi )\otimes
U_{CNOT}(A:a)^{\dagger }\otimes U_A(r)^{\dagger }.
\end{equation}
The effects of decoherence on RSP can be denoted by the state
fidelity between $\rho _B$ and $\rho (\overrightarrow{r})$ (to be
prepared),
\begin{equation}
F\left( \rho (\overrightarrow{r}),\rho _B\right) =Tr[\sqrt{\rho (%
\overrightarrow{r})^{1/2}\rho _B\rho (\overrightarrow{r})^{1/2}}].
\end{equation}
Here, two types of decoherence are considered, depolarizing and
dephasing.

The entangled state after depolarizing is
\begin{equation}
\rho _{AB}^{\prime }(p)=p\left| \Psi ^{-}\right\rangle _{AB\
AB}\left\langle \Psi ^{-}\right| +(1-p)\frac{I_A}2\otimes
\frac{I_B}2,
\end{equation}
which can be considered as a mixture of maximal entanglement and
maximal mixed state. The result state that Bob gets is
\begin{equation}
\rho _B^{\prime }=\frac{1+pr}2\left| \psi \right\rangle
\left\langle \psi \right| +\frac{1-pr}2\left| \psi _{\perp
}\right\rangle \left\langle \psi _{\perp }\right|
\end{equation}
and the state fidelity is
\begin{equation}
F(\rho (\overrightarrow{r}),\rho _B^{\prime })=\frac 12\left( \sqrt{%
(1+r)(1+pr)}+\sqrt{(1-r)(1-pr)}\right)
\end{equation}
Specially, for maximal entanglement shared $p=1$, $F=1$; for pure
states to
be prepared $r=1$,$F=\sqrt{\frac{1+p}2}$; and for maximally-mixed states $%
r=0 $, $F=1$. We found that the fidelity is independent of $\theta
$ and $\phi $ and only related to the quantum channel ($p$) and
$r$.

For dephasing decoherence,
\begin{equation}
\rho _{AB}^{\prime \prime }(p)=p\left| \Psi ^{-}\right\rangle
_{AB\ AB}\left\langle \Psi ^{-}\right| +\frac{1-p}2\left( \left|
0\right\rangle _{A\ A}\left\langle 0\right| \otimes \left|
1\right\rangle _{B\ B}\left\langle 1\right| +\left| 1\right\rangle
_{A\ A}\left\langle 1\right| \otimes \left| 0\right\rangle _{B\
B}\left\langle 0\right| \right).
\end{equation}
It can be considered as a mixture of maximal entanglement and
classical correlation. The result state and the fidelity of RSP
are
\begin{equation}
\rho _B^{\prime \prime }=\frac{1+p}2\left( \frac{1+r}2\left| \psi
\right\rangle \left\langle \psi \right| +\frac{1-r}2\left| \psi
_{\perp }\right\rangle \left\langle \psi _{\perp }\right| \right)
+\frac{1-p}2\left( \frac{1+r}2\left| \psi ^{\prime }\right\rangle
\left\langle \psi ^{\prime }\right| +\frac{1-r}2\left| \psi
_{\perp }^{\prime }\right\rangle \left\langle \psi _{\perp
}^{\prime }\right| \right)
\end{equation}
and
\begin{equation}
F(\rho (\overrightarrow{r}),\rho _B^{\prime \prime })=\sqrt{\left( \frac{%
\alpha +\beta }2\right) +\sqrt{\left( \frac{\alpha -\beta
}2\right) ^2+\gamma ^2}}+\sqrt{\left( \frac{\alpha +\beta
}2\right) -\sqrt{\left( \frac{\alpha -\beta }2\right) ^2+\gamma
^2}}  \label{101}
\end{equation}
where
\[\left| \psi ^{\prime }\right\rangle =\cos \frac \theta 2\left|
0\right\rangle -e^{i\phi }\sin \frac \theta 2\left| 1\right\rangle
,\left| \psi _{\perp }^{\prime }\right\rangle =e^{i\phi }\sin
\frac \theta 2\left| 0\right\rangle +\cos \frac \theta 2\left|
1\right\rangle ,\]
\begin{eqnarray*}
\alpha  &=&\frac 18\left( \left( 1+p\right) \left( 1+r\right)
^2+\left(
1-p\right) \left( 1+r\right) \left( 1+r\cos 2\theta \right) \right)  \\
\beta  &=&\frac 18\left( \left( 1+p\right) \left( 1-r\right)
^2+\left(
1-p\right) \left( 1-r\right) \left( 1-r\cos 2\theta \right) \right)  \\
\gamma  &=&\frac 18r\left( 1-p\right) \sqrt{1-r^2}\sin 2\theta .
\end{eqnarray*}
Specially, for maximal entanglement shared $p=1$, $F=1$; for pure
states $r=1 $, $F=\frac 12\sqrt{3+p+(1-p)\cos 2\theta }$; and for
$r=0$, $F=1$.

Here the fidelity for depolarizing is found to be independent on
states to
be prepared. While for dephasing, it depends only on the proportion of $%
\left| 0\right\rangle $ and $\left| 1\right\rangle $,
\textit{i.e.} depending on $\theta $, but not on the relative
phase $\phi $. And, obviously, the later is better than former,
which means classical correlation can help remote state
preparation.

\section{Experiments of remote state preparation for single photons}

In this section, an experiment of remote state preparation for
single photons is carried out by using spontaneous parametric down
conversion (SPDC) and linear optical elements. Here, we only
discuss the RSP in dephasing noisy channel. The setup is
represented in Fig.$3$. A pulse of ultraviolet (UV) light pass
through a BBO crystal ($0.5$ mm, cut for type-II phase match) .
The UV pulse is frequency-doubled pulse (less than $200$ fs with
$82$ MHz repetition and $390$ nm center-wavelength) from a
mode-locked Ti: sapphire laser (Tsunami by Spectra-Physics) .
Because of the birefringence of ordinary light ($o$ light) and
extra-ordinary light ($e$ light) in BBO crystal, the state of the
biphoton from SPDC process is no longer the maximal entangled one,
but is the the state like
\begin{equation}
\rho _{AB}(p)=p\left| \Psi ^{-}\right\rangle _{AB\ AB}\left\langle
\Psi ^{-}\right| +\frac{1-p}2\left( \left| H\right\rangle _{A\
A}\left\langle H\right| \otimes \left| V\right\rangle _{B\
B}\left\langle V\right| +\left| V\right\rangle _{A\ A}\left\langle
V\right| \otimes \left| H\right\rangle _{B\ B}\left\langle
H\right| \right) .
\end{equation}
Where $p$ can be adjusted by quartz plate with different thickness.
In our experiment, we only chose $%
p=0.9$ and $0.7$. When $p=0.9$, we use the technology of
tomography to reconstruct density matrix of the state from SPDC
process\cite{Jam01},
\[
\left(
\begin{array}{llll}
0.001875 & -0.018531+0.013719i & 0.002594+0.017125i & 0.01-0.015437i \\
-0.018531+0.013719i & 0.50125 & -0.435688+0.002406i & -0.007469+0.007281i \\
0.002594+0.017125i & -0.435688+0.002406i & 0.494375 & -0.007281+0.005813i \\
0.01+0.015438i & -0.007469+0.007281i & -0.007281+0.005813i &
0.0025
\end{array}
\right) .
\]
The fidelity between $\rho _{AB}(0.9)$ and the upper state we get experimently is $%
99.7\%$. And also, we get the state from SPDC process when
$p=0.7$, the fidelity is $99.5\%$. We have possessed the perfect
quantum channel (noisy entanglement), and next, we can carry out
our schemes of remote state preparation.

In Fig.3, photon A passes through path $2$. First, it is rotated
by QWP1 and HWP1 which realize the unitary operator $U_A(\theta
,\phi )^{\dagger }$(see eq.(5)). Second, photon A transmits a
$3.0$ mm $\beta $-BBO crystal, Whose $o$ axis is fixed
horizontally and $e$ axis is fixed vertically. Here we chose the
time degrees of freedom of photon A passing through BBO crystal to
be the auxiliary qubit ($\left| t_o\right\rangle $ for ordinary
light and $\left| t_e\right\rangle $ for extra-ordinary light).
After $3.0$ mm BBO crystal, the separation of wavepackets between
H$(o)$- and V$(e)$-polarized light is about $350$ $\mu m$. Because
the coherent length of the wavepacket is about 150 $\mu m$ (4 nm
FWHM interference filter is inserted before each
detector), There is  is no superposed parts between the H$(o)$- and V$(e)$%
-polarized light. So a controlled-NOT operation is accomplished
after the photon A passes through the BBO crystal
\[
\left( a\left| H\right\rangle +b\left| V\right\rangle \right)
\left| t_0\right\rangle \rightarrow a\left| H\right\rangle \left|
t_o\right\rangle +b\left| V\right\rangle \left| t_e\right\rangle ,
\]
where the polarization, $\left| H\right\rangle $($\left|
0\right\rangle $)/$\left| V\right\rangle $($\left| 1\right\rangle
$), is the control qubit and the time, $\left| t_o\right\rangle
$($\left| 0\right\rangle $)/$\left| t_e\right\rangle $($\left|
1\right\rangle $), is the target qubit. In our experiment, the
input target qubit is fixed in $\left| t_o\right\rangle $($\left|
0\right\rangle $). Following $3.0$ mm BBO, photon A is operated by
HWP2 which realizes the unitary $U_A(r)$(see eq.(9)). The time
qubit is traced when the photon is detected by a single photon
detector. Photon B passes through path 1. Any single-qubit mixed
states Alice want to prepare can be remotely prepared in Bob's
laboratory by adjusting the angles of QWP1, HWP1 and
HWP2\cite{kwia} at Alice's side, and Bob can use technology of
quantum state tomography to reconstruct the density matrix of it.

To prepare pure states, we fixed the angle of HWP2 $0$ ($r=1$).
QWP1 and HWP1 can help Alice remotely prepare any pure states in
Bob's laboratory. Bob use technology of tomography to reconstruct
the state matrix of it again.

In our experiment, several pre-agreed state sets are selected (see
in Fig.1). For each pre-agreed state set, we finish the process of
RSP when $p=0.9$ and $0.7$ respectively as shown in Fig.4 and
Fig.5. Three sets of pure states are remotely prepared as show in
Fig 4.: 1)the polar great circle on the sphere crossed by X-Z
plane ($r=1,\theta \in [0,\pi ]$, $\phi =0$), 2)the polar great
circle by Y-Z plane ($r=1,\theta \in [0,\pi ]$, $\phi =\pi /2$),
and 3)the equatorial great circle by X-Y plane ($r=1,\theta =\pi
/2,\phi \in [0,2\pi ]$ ). It is found that the fidelity of RSP is
independent of the relative phase $\phi $ and depend on $\theta $
and the noise of entanglement, as eq.(20) shows. Four sets of
mixed states are shown in Fig.5a and 5b: 4)a small circle on the
X-Z plane ($r=\cos ^2\frac \pi 8$, $\theta \in [0,\pi ]$, $\phi
=0$); 5)the zero vector $\overrightarrow{O}$; and two lines on the
X-Z plane: 6) ($r\in [-1,1]$, $\theta =\pi /4$, $\phi
=0$)\cite{Note} and 7) ($r\in [-1,1],\theta =\pi /2,\phi
=0$)\cite{Note}. Fig.5 tell us that the the fidelity of RSP of
mixed state is the cosine function of $\theta $, and the fidelity
of the maximally mixed state is always one(Fig.5a). the effect of
noise to pure states is greater than mixed states (Fig.5b). In
Fig.4 and Fig.5, the solid lines are the theoretical results given
by eq.(20). For example, in Fig.4b, the state from the polar great
circle on the sphere crossed by X-Z plane ($r=1,\theta \in [0,\pi
]$, $\phi =0$) are remotely prepared and the parameter of channel
$p=0.7$, So the fidelity curve given by eq.(20) is $F=\frac
12\sqrt{3.7+0.3\cos 2\theta }$. It is found that all experiment
results (the square dots) match the theoretical prediction (the
solid lines) well. The imperfection of our experimental results
only comes from the fluctuation of the coincidence counts and the
limited precision ($2^{\circ }$) of waveplates (HWPs and
QWPs)\cite{Jam01}.

\section{Conclusion}

In conclusion, we present a practical and general scheme of remote
preparation for pure and mixed states. An auxiliary qubit and
controlled-NOT operation are used in the scheme. The effects of
noisy entanglement are discussed for two important types of
decoherence, depolarizing and dephasing, by the state fidelity.
The fidelity for depolarizing is found to be independent of
$\theta $ and $\phi $ of  states to be prepared and only related
to the quantum channel ($p$) and
$r$. While for dephasing, it depends only on the proportion of $%
\left| 0\right\rangle $ and $\left| 1\right\rangle $,
\textit{i.e.} depending on $\theta $, but not on the relative
phase $\phi $. And the dephasing entanglement is always better
than the depolarizing one for RSP, which implies classical
correlation is \ helpful for RSP. In our experiment, we
successfully complete RSP of pure and mixed states via dephasing
entanglement by using spontaneous parametric down conversion
(SPDC) and linear optical elements. Our scheme can be carried out
remotely by using optic fiber or in free space.

\begin{center}
\textbf{Acknowledgment}
\end{center}

We wish to thank Yong-Sheng Zhang, Guo-Ping Guo and Ming-Yong Ye
for interesting and helpful discussion. This work was funded by
the National Fundamental Research Program (2001CB309300), the
Innovation Funds from Chinese Academy of Sciences, and also by the
outstanding Ph. D thesis award and the CAS's talented scientist
award rewarded to Lu-Ming Duan.

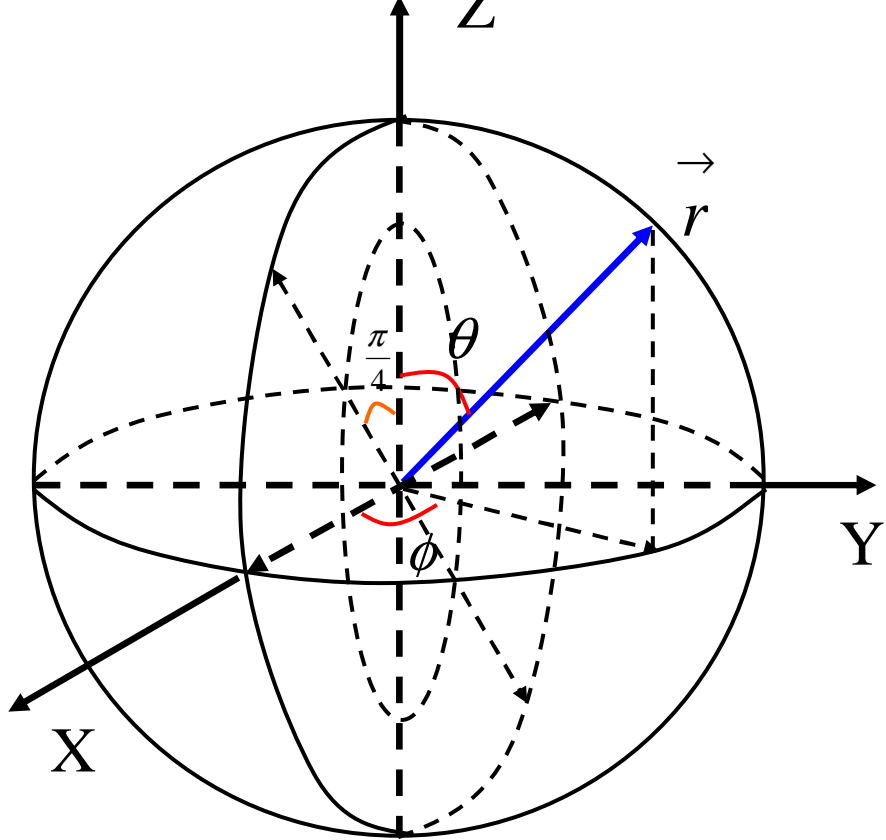
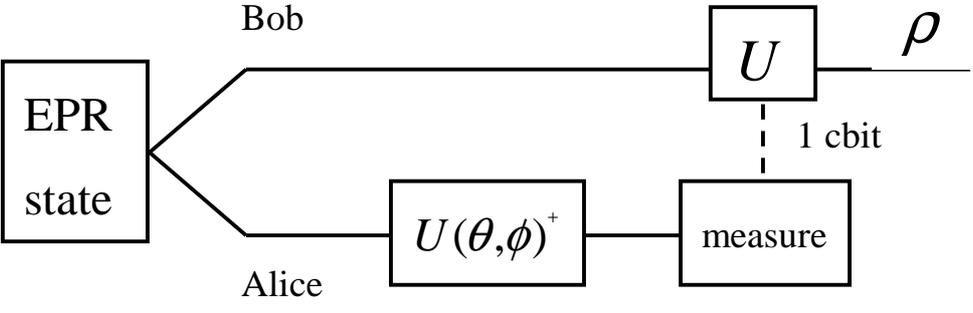
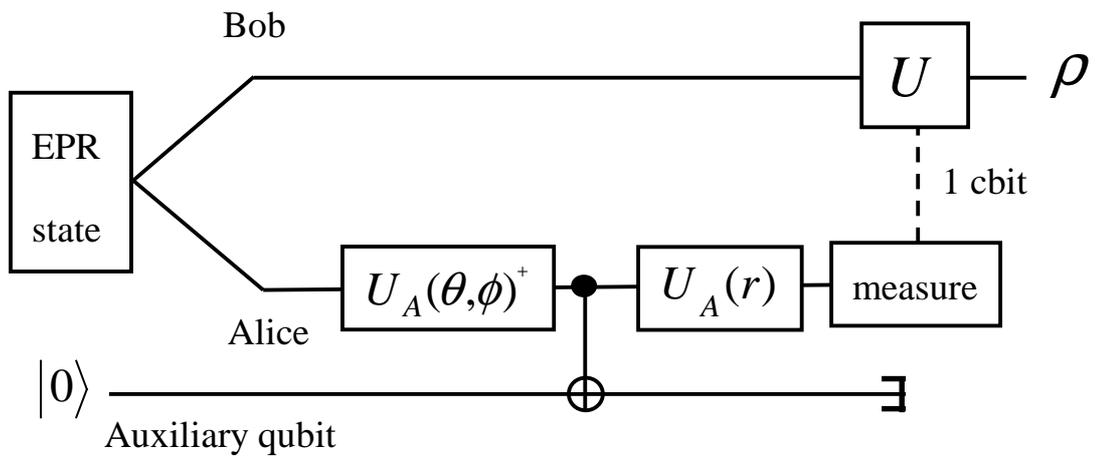
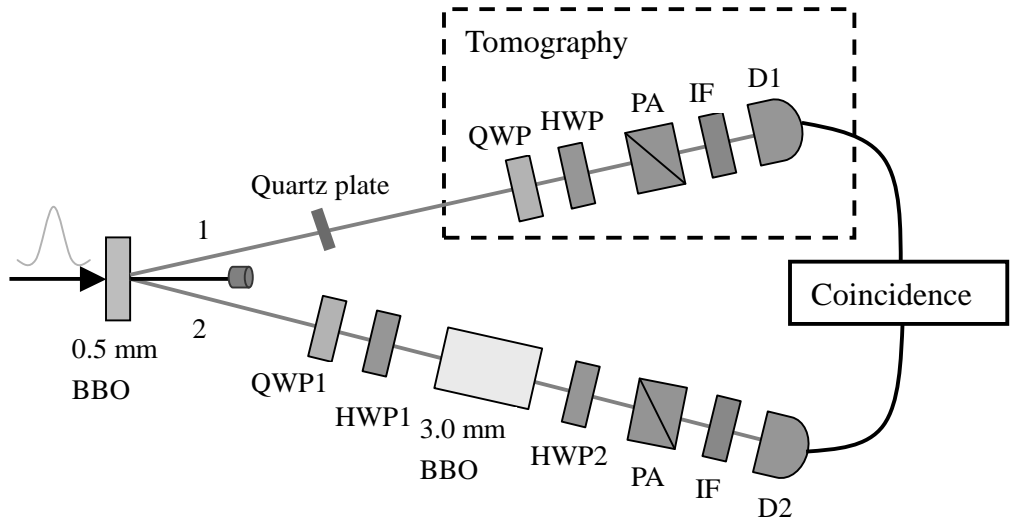
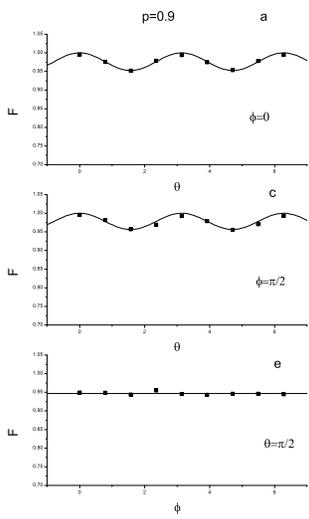
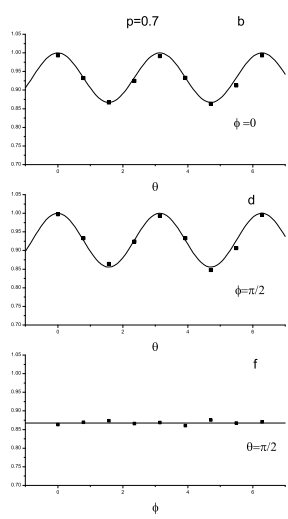
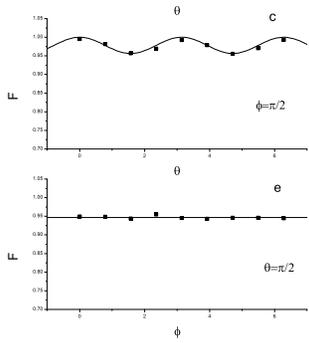
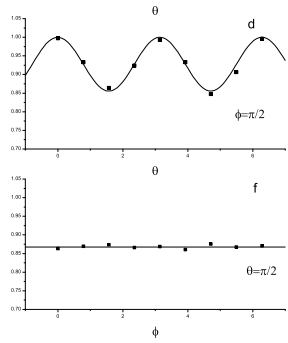
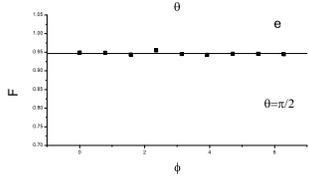
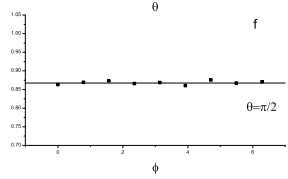
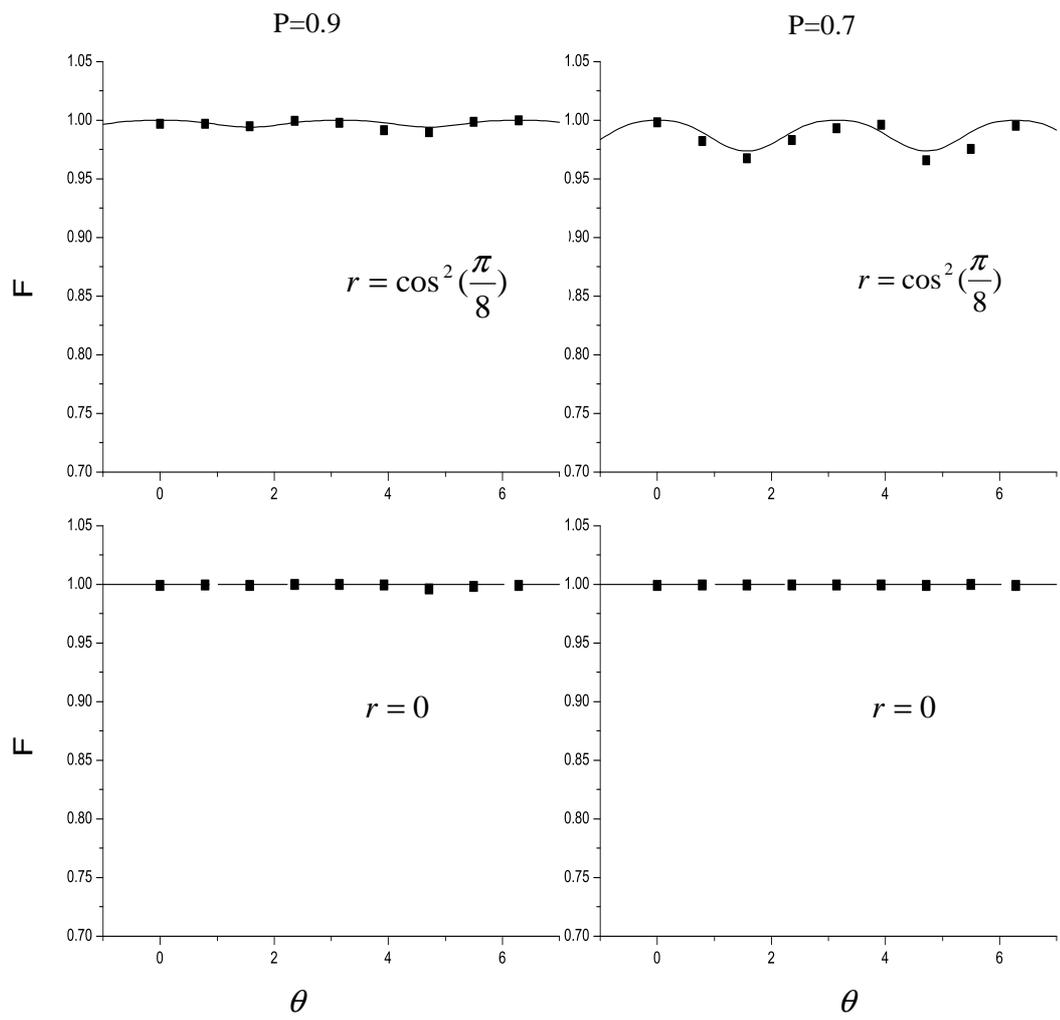
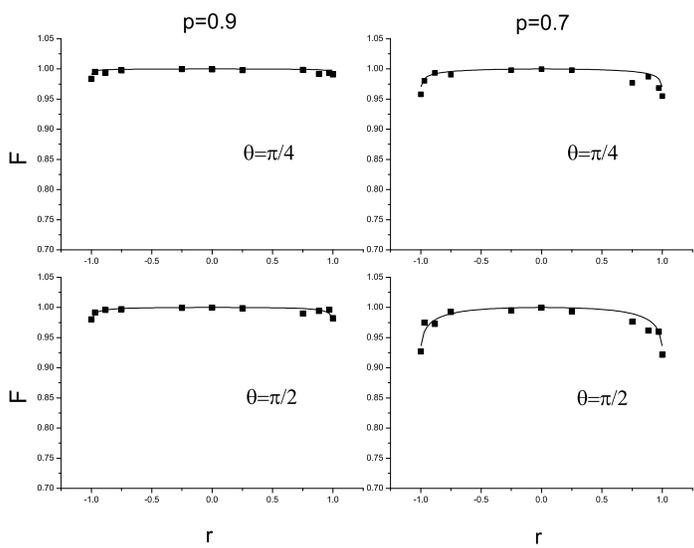
\end{document}